\documentstyle[twocolumn,aps,prl,epsf]{revtex}

\begin{document}

\title{{\bf Folding and Stretching in a Go-like Model of Titin}}

\author{{\bf Marek Cieplak$^{1,2}$, Trinh Xuan Hoang$^3$, and
Mark O. Robbins$^1$}}

\address{
$^1$ Department of Physics and Astronomy, The Johns Hopkins University,
Baltimore, MD 21218\\
$^2$ Institute of Physics, Polish Academy of Sciences,
Al. Lotnik\'ow 32/46, 02-668 Warsaw, Poland \\
$^3$ International School for Advanced Studies (SISSA),
via Beirut 2-4, 34014 Trieste, Italy}

\maketitle

\vskip 40pt
\noindent
$^*$Correspondence to: \\
Marek Cieplak,\\
Institute of Physics, \\
Polish Academy of Sciences, \\
Al. Lotnik\'ow 32-46  \\
02-668 Warsaw, Poland\\
Tel:  48-22-843-7001\\
Fax:  48-22-843-0926\\
E-mail: mc@ifpan.edu.pl

\vskip 40pt
\noindent
Grant sponsors: NSF DMR-0083286, TIPAC (Johns Hopkins), 
and KBN (Poland) -  2P03B-146-18.

\vskip 40pt
\noindent {\bf
Keywords: mechanical stretching of proteins; protein folding; Go model;
molecular dynamics; atomic force microscopy; titin}

\newpage
\begin{abstract}
{Mechanical stretching of the I27 domain of titin and of its
double and triple repeats are
studied through molecular dynamics simulations
of a Go-like model with Lennard-Jones contact interactions.
We provide a thorough characterization of the system and correlate the
sequencing of the folding and unraveling events with each other
and with the contact order. The roles of cantilever stiffness
and pulling rate are studied. Unraveling of tandem titin structures
has a serial nature. The force-displacement curves in this coarse-grained
model are similar to those obtained through all atom calculations.
}
\end{abstract}


\section*{INTRODUCTION}
The giant protein molecule titin, also known as connectin, is responsible
for the elastic and extension properties of smooth, skeletal,
and cardiac muscles. \cite{Labeit0,structure,Ericson,Schulten}
Titin spans half of the sarcomere, the repeat segment in muscle
fibers, and has been 
implicated in certain heart diseases.
The biological importance of titin and the nature of its function
make it an ideal candidate for studies of the mechanical properties
of proteins.\\

Titin consists of about 30 000 amino acids which are organized into
about 300 domains that form the so-called A-band followed by the I-band.
Domains which are similar to fibronectin type III are located only in the
A band, whereas immunoglobulin-like (Ig) domains appear throughout
the length of titin. In the middle of the I-band there is
one special
domain \cite{Labeit} called PEVK which consists of between 163
(in a cardiac titin), and 2174 (in a skeletal titin) residues.
The number of Ig domains in the I-band ranges from 37 in cardiac
titin to 90 in skeletal titin. The native structure of only one of these Ig
domains, I27 (which is especially stable), 
has been resolved through NMR spectroscopy \cite{structure} 
and found to have the architecture of a $\beta$-sandwich.  
The remaining domains are believed to be similar in structure, even though
the sequence similarity is only of order 25\%.\\

The first mechanical studies of titin concluded that small stretching
forces affect primarily the PEVK domain \cite{Ericson,Linke}
and still larger forces induce extension of the Ig domains \cite{Granzier}.
Recent experiments on stretching of single titin molecules by optical tweezers
\cite{Simmons,Bustamante} and atomic force microscopy \cite{Gaub} showed
relevance of both PEVK and Ig domains to the mechanics of titin.
The overall picture is that the I-band accommodates stretch by straightening of 
the Ig domains and by unfolding of the PEVK domain.
Unraveling  of the 25 consecutive Ig domains
resulted in a train of sawtooth force patterns as a function of tip
displacement that repeated in a periodic manner \cite{Gaub} which made
Ericson \cite{Ericson} call titin a ``weird spring.''
Initiation of
stretching of the Ig domains was found to require a very high force sustained
for only a short tip displacement
\cite{Gaub}. The force
drops once the domain is destabilized, and later force peaks are not as high.\\ 

Natural titins are inhomogeneous. However, protein engineering has 
recently allowed production of tandem repeats of identical Ig 
modules \cite{Marszalek0,Marszalek,Fernandez}.
Studies of proteins
made of 8 and 12 Ig domains have indicated \cite{Marszalek1} that
the hydrogen bonds allow the domains of titin to stretch reversibly
only up to some limiting extension.
Beyond that limit, misfolding takes place.
This is consistent with the studies of mechanical unfolding of secondary
structures in the companion paper \cite{first} where proteins refolded
to the proper sequence until an irreversibility length was exceeded.\\

Understanding of the unraveling of titin has been facilitated by 
all atom (no water molecules) computer 
simulations \cite{Schulten,Zhang,Paci1,Paci2},
which indicated existence of a large bottleneck to unfolding of a single
domain at small
end-to-end extensions and pointed to a serial character of the 
many-domain unraveling in titin \cite{Paci2}.
Klimov and Thirumalai have considered simplified coarse-grained
lattice \cite{Thirumalai2} and off-lattice \cite{Klimov} models.
The latter were based on a model \cite{Veitshans} which contained
three kinds of amino acids and Lennard-Jones interactions
between pairs of hydrophobic residues.
Two model sequences with four-stranded $\beta$-barrel topology
were considered. Klimov and Thirumalai \cite{Klimov}
noted that thermal unfolding
appeared to proceed along pathways which were distinct from the
stretching trajectories.
These studies led them
to suggest
that a natural way to
characterize constant force-induced unfolding is in terms  of a phase 
diagram on a plane defined by the force and the concentration of denatured
fragments.\\

In this paper, we also consider coarse-grained models since
these models allow for a thorough 
characterization of mechanical, equilibrium and
folding properties which is essential to an understanding
of the system. Specifically, we extend our
analysis of the unfolding of secondary structures \cite{first}
to  Go-like models \cite{Goabe} of one, two, and three Ig domains.
These models are defined through the experimentally determined native
structures, and they capture essential aspects of the important role played
by the native geometry on the folding mechanism
\cite{Stakada,Micheletti,Amos}.
We first focus on a single domain and examine its folding characteristics.
We determine the characteristic folding time as a function
of temperature,  
establish the succession of folding events, and relate
it to the contact order -- the separation of two amino acids
along the chain. We then determine the succession of events
in mechanical unraveling and again relate it to the contact order.
We explore cross-correlations between the thermal and mechanical
event sequencing, and single out long-ranged contacts as 
providing a link between the two. 
The protein is stretched by attaching its ends to harmonic
springs to model a Hookean cantilever.
One end of the cantilever is displaced with a constant speed $v_p$.
We study the dependence of the force - displacement curves
on cantilever stiffness and pulling rate.
Finally, we study tandem arrangements of several
domains and show that they unravel in a serial manner that is
in sharp contrast to the parallel unraveling found for
two $\alpha$-helices in tandem \cite{first}.\\

\vspace*{0.5cm}

\section*{MODEL AND METHOD}

The Go-like \cite{Goabe,Stakada} coarse-grained model we use
is explained in references \cite{Hoang,Hoang1,optimal}.
The native structure of I27 is taken from the
PDB \cite{PDB} data bank where  it is stored under the name 1tit
which we shall use as an alternative to I27. The coarse-grained 
picture of 1tit is shown in Figure 1. 1tit consists of 98 residues
which are organized into eight $\beta$-strands and connecting turns.
There are no $\alpha$-helices in 1tit.
In the Go model, amino acids are represented by point particles,
or beads, which are located at the positions of the
C$\alpha$ atoms.
Consecutive beads in the chain are tethered by
anharmonic forces.
The potentials between the beads
are chosen so that the native structure minimizes
the energy.
The interaction between beads that form native contacts
(defined as C$\alpha$ atoms separated by less than 7.5 \AA)
are of the Lennard-Jones type whereas the interactions
for non-native contacts are
purely repulsive. The Lennard-Jones couplings are scaled by a
uniform energy parameter $\epsilon$ and the characteristic lengths,
$\sigma _{ij}$ are contact dependent.
Tandem structures of two or three domains
are constructed by repeating 1tit domains in series
with one extra peptide link between the domains.\\

The model of the pulling cantilever is as in the preceding 
paper \cite{first}. 
Both ends of the protein are attached to harmonic springs
of spring constant $k$.
The outer end of one spring is held fixed, and the 
outer end of the other is pulled at constant speed $v_p$.
The stretching is implemented parallel to the initial
end-to-end position vector.
This corresponds to stretching the protein with a cantilever
of stiffness $k/2$ at a constant rate.
The net force acting on the bead attached to the moving end
is denoted by $F$, the cantilever displacement
is denoted by $d$, and the end-to-end distance of the protein by $L$.
The case of constant force was also considered, but has an all-or-nothing
character that yields little information.\\

Two cantilever stiffnesses and two velocities are considered.
The case of stiff cantilevers corresponds to
$k = 30 \epsilon / $\AA$ ^2$, and 
$k\;=\;0.12 \epsilon / $\AA $^2$ for soft cantilevers.
The case of slow pulling corresponds to a cantilever velocity
of 0.005 \AA/$\tau$.
Here $\tau =\sqrt{m \sigma^2 / \epsilon} \approx 3$ps
is the characteristic time
for the Lennard-Jones potentials, where $\sigma=5$\AA\ is a typical value
of $\sigma_{ij}$ and $m$ is the average mass of the amino acids.
Results begin to become rate dependent just below the
fast pulling rate of $v_p= 0.5 $\AA/$\tau$.\\

To control the temperature, $T$, of the system and mimic the effect of solvent
molecules, the equations of motion for each bead include Langevin
noise and damping terms.\cite{Grest}
A damping constant of
$\gamma \; = \; 2 m/\tau$ is used, where $m$ is the mass
of the bead.
It has been argued \cite{Veitshans} that realistic values
of the solvent damping are 25 times larger.
However, using a smaller damping rate decreases the required
simulation time without affecting the sequencings of
events.\cite{Hoang,Hoang1,thir}
Almost all data correspond to using the same average mass for all
amino acids,
but we also studied the effect 
of amino acid dependent masses on folding times.\\

All folding times
were determined by considering the median times, over 201 trajectories,
needed to form all native contacts. The criterion for forming
a contact is that the distance between the corresponding beads is less than
1.5$\sigma _{ij}$.
As explained in the previous paper \cite{first},
our studies of the mechanics of stretching are performed at $T$=0 in
order to eliminate the need for extensive averaging over trajectories
and to get signals that depend primarily on the energy landscape.
The precise role of the temperature on the force-extension curves 
remains to be elucidated.\\

\vspace*{0.5cm}

\section*{RESULTS AND DISCUSSION}

\noindent
\underline{\bf Folding properties of 1tit}

The sequencing of folding events depends on temperature
and becomes smoothest and most natural at the temperature
of fastest folding $T_{min}$ \cite{Hoang1}. 
For most proteins, the characteristic folding time $t_{fold}$ rises
rapidly on either side of $T_{min}$.
Figure 2 shows that the dependence of
$t_{fold}$ on $T$ for 1tit
exhibits an unusually broad basin of optimality that extends
from about 0.2 to 0.5 $\epsilon /k_B$. 
Another important temperature is the folding temperature, $T_f$,
at which the equilibrium probability, $P_0$, 
of finding the system in its native conformation is one half.
Figure 3 shows $P_0$ vs. $T$ as determined 
from $\sim$ 20 trajectories that lasted for 60 000$\tau$ each.
The value of $T_f$ $=0.26 \epsilon /k_B$
is in the basin of optimal folding times, implying that
the system is a good folder. This is also confirmed by studies
of the $T$-dependence of the specific heat $C_v$ and structural 
susceptibility per bead $\chi _s$ as defined by  Camacho and
Thirumalai \cite{Camacho}. Figure 3 shows that
the peaks in our calculated $C_v$ and $\chi _s$ almost
coincide.
This coincidence has been identified
as another signature of good foldability \cite{Klimov2}.\\ 

The values of the masses of the amino acids affect the equations
of motion for the individual beads. Figure 2 demonstrates that
taking into account the amino acid dependent masses of the beads
does not affect the folding times in any noticeable manner.
This is consistent with the overall coarse-grained character
of the Go model. In more realistic models, however, the non-uniformity
of the masses, and, more importantly, the non-uniformity of the
amino acid shapes and chemical functions,
are expected to have an impact on the
kinetics of folding.\\

The broad minimum in folding time makes it difficult to define a
precise value of $T_{min}$.
We chose to study the sequencing of folding events at
$T=0.25 \epsilon /k_B$ since this temperature is
in the bottom of the basin of optimal folding times and 
close to the folding temperature $T_f=0.26\epsilon /k_B$.
The folding process at this $T$ is characterized in two ways.
The first, shown in Figure 4, is a plot of the average time to establish
a native contact vs. the contact order $|j-i|$, defined as the distance
along the backbone between amino acid $j$ and amino acid $i$.
The second characterization, shown in Figure 5, is through a plot 
of the contact matrix which indicates which beads make a contact.\\

Figure 4 shows that folding of I27 takes place in stages separated by
substantial time gaps.
All short range contacts ($|j-i| < \tilde 20$)
are established within the first 600 $\tau$.
In the next stage, occurring between 1400 and 1700 $\tau$, 
most of the intermediate and some of the long range contacts form.
Then, around 2100 $\tau$,
most of the longest-ranged contacts are established. This is
followed by the completion of the structure through building
up of the remaining intermediate-range contacts.
Thus the sequencing of the folding events 
takes place in stages that are governed, to a large extent,
by the contact order. The latter conclusion is consistent with the findings
in references \cite{Unger,Chan,Plaxco,Ruczinski}.
In addition to the 217 contacts
shown in Figure 4, there are also 87 contacts with
$|j-i|=2$. These contacts of shortest range are established
rapidly and are not displayed for the sake of clarity.\\

The kinetics of folding events can also be gleaned from
the contact map shown in Figure 5.  Here, the symbols indicate
the stage at which a given contact is formed. This representation
allows one to infer details of the secondary structure formation
that are only implicit in the index that defines the contact order.
The last to form are the intermediate-ranged antiparallel
$\beta$-sheets which cannot be established before
contacts of the longest range lock the overall topology in place.\\

In the notation of Figure 1, the average folding trajectory
proceeds according to the scenario:  
$\beta_7 + \beta_8 \rightarrow \beta _2 + \beta _3
\rightarrow \beta _5 + \beta _6 \rightarrow \beta _1 + \beta _3
\rightarrow \beta _3 + \beta _6 \rightarrow \beta _2 + \beta _8
\rightarrow \beta _1 + \beta _8 \rightarrow \beta _4 + \beta _7$, 
i.e. first $\beta _7$ connects with $\beta _8$, then
then $\beta _2$ with $\beta _3$ and so on.
Although this average pattern is the most common,
it was followed by only 24 out of 100 individual trajectories.
In contrast, more than half of the individual trajectories agreed
with the average succession
for the proteins studied in refs. \cite{Hoang,Hoang1} 
(crambin, CI2 and the SH3 domain). 
This difference is related to the longer sequence length of 1tit
and the greater number of events needed to fold it to the native
state.
Other common trajectories correspond to permutations in contact making.
For instance, 15 trajectories have the same sequence of events
as the average succession except that the order of
$\beta _2 + \beta _3$ and $\beta _7 + \beta _8$ is reversed.
In 45 trajectories the last four stages are identical to the
average succession. 
A total of 90 trajectories establish $\beta$-sheets with
low contact order ($\beta _7 + \beta _8$, $\beta _2 + \beta _3$,
$\beta _5 + \beta _6$, $\beta _1 + \beta _3$) before
the sheets with high contact order are formed ($\beta _3 + \beta _6$,
$\beta _2 + \beta _8$, $\beta _1 + \beta _8$, $\beta _4 + \beta _7$).\\

\vspace*{0.5cm}
\noindent
\underline{\bf Stretching of 1tit}

The two snapshots of the mechanically unfolded model of 1tit shown in 
Figure 6 indicate that stretching affects the short and long range
contacts simultaneously.
Both ends of the protein straighten over
longer and longer length scales but the central "knot" gets
unraveled as well, starting first at the longest-ranged contacts that 
pin the structure.
These longest ranged contacts are {\em not} those that were
established at the last stage of thermal folding but they
do arise towards the end of folding.
Before we look into the
issues of event sequencing in more detail, we discuss the 
force-extension curves.\\

Figure 7 shows the force as a function of cantilever displacement
for the two values of stiffness.
The curves are terminated when the protein is fully extended and
the sharp rise in force at the end of the curves reflects stretching
of covalent bonds along the backbone.
As in the preceding paper, the force curves show a series of upward
ramps followed by rapid drops where contacts break.
The slope of the upward ramps is the combined stiffness of the protein
and cantilever.
The protein is softer than the stiff spring and its elastic properties
dominate in this case.
The opposite applies for soft springs, and the slope of the upward ramps
nearly coincides with $k/2=0.06\epsilon/$\AA$^2$.
The ramps end when one or more contacts break.
This allows the protein to extend and the force drops.
As the cantilever stiffness decreases, the force drops more slowly
with increases in the length $L$ of the protein.
If the extension due to breaking one set of contacts is not
large enough, the force may remain above the threshold for breaking
the next set of contacts.
This leads to large avalanches where many contacts break in a single
extended event.
The stiff spring is able to resolve nearly all independent contact
ruptures, while they coalesce into a much smaller number of large
events in the soft spring case.
Increasing the velocity 100-fold (Figure 7, dotted lines), 
limits the ability of tension to equilibrate along the chain
and causes further merging of events.\\

For all force curves, the largest maximum occurs near the beginning
of stretching.
This peak represents the main bottleneck to mechanical unfolding.
Subsequent peaks are visible, but are less than half as large.
A few small drops are also visible on the way up to the main peak
at which a total of 28 contacts break.
Similar curves for mechanical unfolding of secondary structures
showed a very different pattern.
In the cases studied \cite{first}, the force needed to break
bonds tended to increase or remain constant until nearly all
contacts had failed.\\

It is remarkable that room temperature all atom simulations (CHARMM-based)
by Lu et al. \cite{Schulten} for a stiff cantilever
yield a pattern (their Figure 5)
which is very similar to the one shown in the top panel of Figure 7.
Indeed, their simulations place the main "burst" of contact rupture
as occurring between 10 and 17 {\AA}  which is quantitatively
consistent with our results. 
The peak in $F$ in our simulations corresponds to
end-to-end extensions of $d$=17.1{\AA} and 17.3{\AA} for stiff
and soft springs respectively.
Our results also agree qualitatively
with those obtained by Paci and Karplus \cite{Paci2}.
They have performed a controlled biased force 
calculation, without an explicit cantilever but also CHARMM-based, and found
that the large bottleneck to unfolding arises at 
end-to-end extensions of order 6 {\AA}.
The lower extension in this simulation may
be due to the different ensemble used
to implement the pulling force \cite{Paci2}.
Marszalek et al. have obtained a much smaller critical extension of about
2.5 \AA by fitting a two-state model with worm-like-chain elasticity to
experimental results \cite{Marszalek}. 
However, all the simulations described above reveal an energy landscape
with extra minima before the main force peak, and the elasticity is
also more complicated than that of a worm-like-chain.
Both previous all atom simulations\cite{Schulten,Paci2} identified the main
burst with concurrent breaking of six interstrand hydrogen bonds
between $\beta$-strands A' and G located near the C-terminus
(denoted respectively by $\beta _2$ and $\beta _8$ in Figure 1).
This identification is discussed further in Ref. \cite{SchultenLu},
and Figure 9 (see below) shows that the same set of contacts breaks at
the main force peaks in Figure 7.\\

In the subsequent unraveling of 1tit there are further bursts
of contact rupture
but none of them is as significant as the first one.
When the cantilever is displaced by about 300 \AA (the top panel of Figure 7),
the domain is fully stretched and the force starts to increase rapidly
indicating an incipient rupture.
At this value of the displacement the end-to-end distance $L$
is about 342 {\AA} which corresponds to an almost 8-fold stretch
relative to the native value of 43.19 \AA.
The experimental data show the immunoglobulin domain unraveling
on extension from 40 to 300 {\AA} \cite{Soteriou,Erickson1}
which is consistent with the range in our model.\\

Figure 8 shows the number of native contacts, $n_{\small NAT}$,
and the energy, $E$, of the model protein as a function of cantilever
displacement at small pulling rates. For the stiff cantilever
the dependence on $d$ appears to be nearly continuous and monotonic, but 
a closer inspection reveals the presence of small jumps at certain
values of $d$.
These correlate with the bursts in the $F$ vs. $d$ curve at the top
of Figure 7.
For the soft cantilever the steps involve much larger changes
in both $n_{\small NAT}$ and $E$, and the
synchronization with drops in the force curve is more evident.\\

To illustrate the unfolding sequence, we first plot the cantilever
displacement where each bond opens, $d_u$, as a function of
contact order.
Results for stiff and soft springs are shown in Figs. 9 and 10,
respectively.
Note the presence of clear clusters in the $d_u$ vs. $|j-i|$ plane.
The same set of bonds are clustered in the stiff and soft plots,
however the shapes of the clusters are different.
The more horizontal character of the clustering in the soft case
is due to coalescence of multiple bonds into coherent breaking events. \\

Several differences between mechanical unfolding and thermal folding
are evident when one compares Figures 9 and 10 with Figure 4.
First, in thermal folding, the intermediate range ($|j-i|$ of around 40)
and long range contacts are each divided into two time-separated groups.
In contrast,
in mechanical unraveling all long range contacts cluster together
and (except for several outliers) the intermediate contacts
also form a single cluster.
Second, in thermal folding the short range contacts get established 
rapidly whereas in stretching they continue to rupture throughout
almost the entire unfolding process.
Notice though that stretching leads to breaking of nearly all the long-range
bonds before the short-range contacts begin to fail.\\

These differences are also evident when the contact map 
for unfolding with a stiff cantilever shown in Figure 11 is compared
with the thermal folding map shown in Figure 5.
In both cases the symbols indicate the stage at which the event occurred
and local clusters tend to evolve in the same stage.
However, the order of unraveling shows no simple correlation with that
of folding.
This is particularly true for the contacts of short range that
lie along the diagonal in Figures 5 and 11.
All of these bonds form at early times during folding, but they
break in the second and fourth stages of mechanical unfolding.
The intermediate range bonds all unravel in the third stage of
unfolding, but form in the second and fourth stages of unfolding.
Only the long-range bonds act together in both cases,
breaking in the first stage of unfolding and forming in the third
stage of folding.\\

To illustrate the cross-correlation between stretching and folding,
the breaking distance for each contact is plotted against
folding time for stiff
and soft cantilevers in Figures 12 and 13, respectively.
Different symbols are used to indicate different ranges of bond order.
The low order contacts (3-11) span the full vertical range but are confined
to short times. 
This same lack of correlation is seen in the case of two $\alpha$-helices
connected together \cite{first}.
The long-range bonds show a clear anti-correlation, occurring at
short distances and long times.
Intermediate-range bonds are localized in a narrow range of $d_u$,
but clustered into two different time intervals.\\

\vspace*{0.5cm}
\noindent
\underline{\bf Irreversibility length}

Figure 2 suggests that the folding time is infinite at $T$=0, i.e. when
one starts with a typical open conformation it will never find its way
to the native state. However, we find that when one stretches the protein 
slowly by less than some irreversibility length $L_{ir}$ \cite{first},
the protein will fold back after release.
Figure 14 shows the non-monotonic
dependence of refolding time on the end-to-end length $L$ of the protein
at the point of release.
The plots are terminated when the protein begins to misfold.
This limiting length corresponds to $L_{ir}$, and is about 56\AA.
The precise value of $L_{ir}$ depends on the properties of the cantilever,
however the variations in the cantilever displacement
at the onset of irreversibility, $d_{ir}$, are much larger.
For the stiff and soft cases considered here the values of $d_{ir}$ are
12.6 {\AA} and 61.7 {\AA}, respectively.
As for simpler proteins in the companion paper \cite{first}, the onset
of irreversibility is associated with the same set of broken
contacts.  From Figure 7, we see that both values of $d_{ir}$ lie about
half way up the ramp to the first large force peak.  From Figure 8,
we find that 20 contacts are broken at $d_{ir}$
for both cantilever stiffnesses.

\vspace*{0.5cm}
\noindent
\underline{\bf Several domains in tandem} 

Titin consists of many different domains of globular proteins.
Since the structure of one of these domains is known and since the
structure is the primary experimental input to Go modeling,
we consider tandem structures made of repeats of 1tit.
Figure 15 illustrates  unraveling of two domains. It indicates that
the unraveling proceeds basically in series,
as in ref. \cite{Paci2}, whereas unraveling of two
$\alpha$-helices has been found to proceed in parallel \cite{first}.
On closer inspection, it appears that the two domains start
unraveling together but after a small number of contacts get broken
in both domains, only the forward domain continues to unfold and
only after
this process is completed, does the backward domain engage in action.
This is clearly seen in Figure 16  which shows the displacement
at which contacts break vs. their contact order  for the case of the
slowly pulled stiff cantilever.
The data points are marked to differentiate between the forward and
backward domains.
The closed symbols from the backward domain lie almost entirely in
the upper half of the
$d_u$ vs. $|j-i|$ plane and are a nearly perfect repeat of the pattern
formed by the open symbols representing the forward domain.
Simulations with three domains show another periodic repeat
of the same pattern.\\

The serial character of unraveling is also seen from the force-displacement
curves shown in Figures 17 and 18 for two and three domains respectively.
Independent of the cantilever stiffness, there is a nearly periodic
repetition of the events that take place in one domain.
The reason for the serial character of the unraveling is the
existence of the high force peak at the beginning of the unfolding process.
Once this peak is past, a domain unfolds completely at a lower force
which is not sufficient to initiate unfolding of other domains.
Only when the first domain is completely unraveled can the force rise
and initiate unfolding of another domain.\\

Experimental data \cite{Gaub} also show a periodically repeated
sawtooth-like pattern but there is an overall upward trend in the curves
as one unravels successive domains.
The reason for this trend is the fact that the domains in series
are not identical and the most weakly bound 
of them all unravels first. When the domains are made
identical, there is no trend \cite{Marszalek1}.
The serial character of unraveling is also present
in many natural adhesives \cite{Hansma} and
has been described as analogous to the story of Sisyphus
of the Greek mythology: "The case of extending a modular fibre is
analogous. One needs to pull hard, and do work, but before
the breaking point ('the summit') is reached, a domain unfolds or
a loop opens, and the energy stored in the fibre is dissipated as heat.
Then, the fibre has to be pulled on again, until the next domain breaks
and so on." \cite{Hansma}\\

\vspace*{0.5cm}

\section*{CONCLUSIONS}

Force spectroscopy is a useful tool for obtaining information about
the strength of modules in a protein and to infer
relationships between structure and function. However, inferring
information about folding pathways from mechanical data turns out to be
far from straightforward.
The companion paper showed that folding of the simple secondary structures
considered is uniquely related to their mechanical unraveling, but that
the sense of the correlation in $\alpha$-helices is opposite to that
in $\beta$-hairpins.
For the more complicated geometry of titin considered here,
any correlation between stretching and folding
appears to be restricted to the long-ranged contacts.
These contacts tend to 
form last and unravel at the beginning, although how soon
depends on the nature of the cantilever. It would be
useful to study other proteins using similar techniques to determine
possible systematics in behavior. The use of simplified models,
such as the Go-like model considered here, is encouraged
since we have found the mechanical results to be strikingly
similar to those obtained through all atom simulations.
The irreversibility length may be a useful parameter to determine
in experimental studies of mechanical misfolding.

\section*{ACKNOWLEDGMENTS}
We appreciate discussions with J. R. Banavar and T. Woolf 
which motivated parts of this research. This work was funded
by KBN, NSF Grant DMR-0083286 and the Theoretical Interdisciplinary
Physics and Astrophysics Center.




\begin{figure}
\epsfxsize=2.5in
\centerline{\epsffile{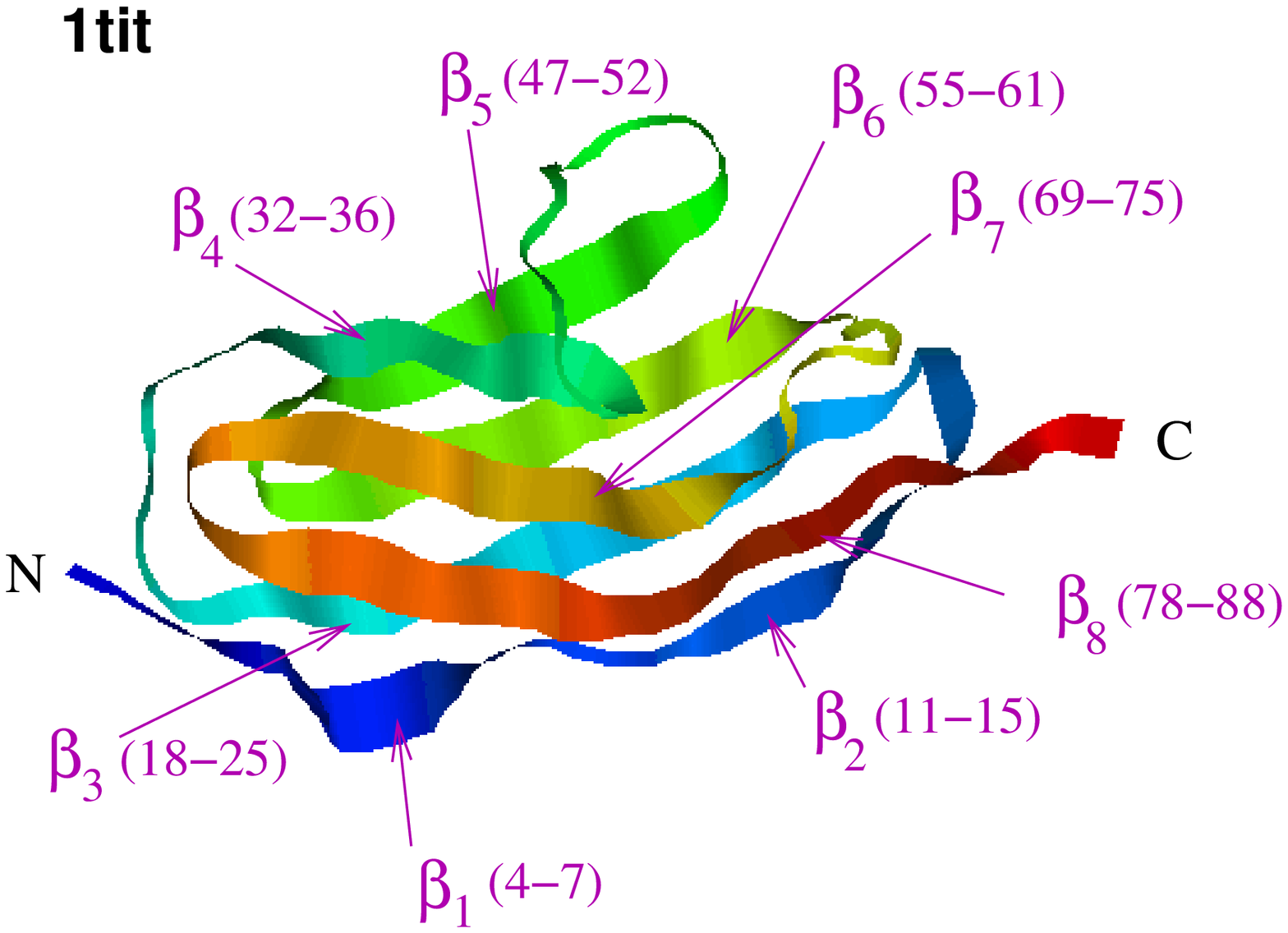}}
\caption{
The ribbon representation of the domain 1tit.
The symbols indicate $\beta$-strands together with the sequence position
of the amino acids involved.
}
\end{figure}

\begin{figure}
\epsfxsize=2.5in
\centerline{\epsffile{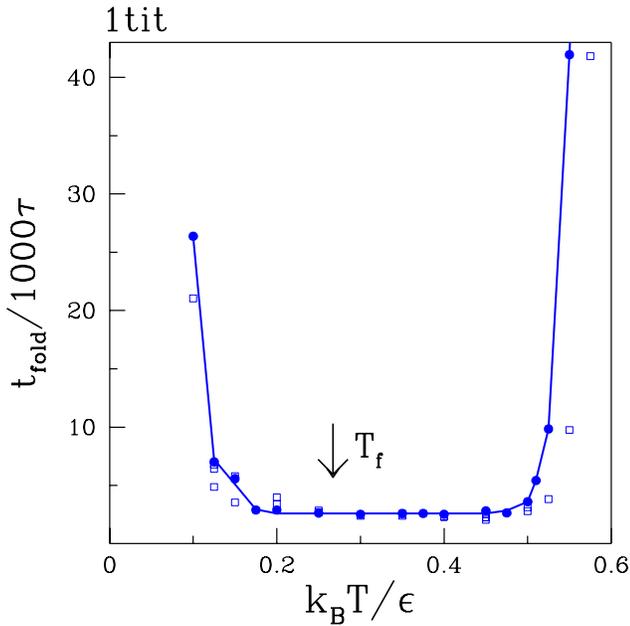}}
\vspace*{2cm}
\caption{
Median folding time for the Go-like model of 1tit.
The solid symbols correspond to simulations with uniform masses and
the open squares to those using
the actual masses of the amino acids in the sequence. The arrow indicates
the value of the folding temperature.
}
\end{figure}

\begin{figure}
\epsfxsize=2.5in
\centerline{\epsffile{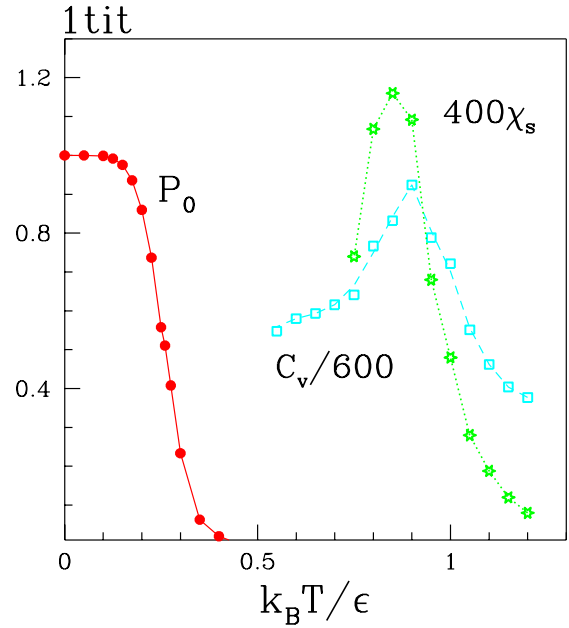}}
\vspace*{2cm}
\caption{
Equilibrium properties of 1tit. The circles show the
probability for the protein to be in the native state,
the stars show the dimensionless structural
susceptibility, and the squares show the specific heat in
Lennard-Jones units.
}
\end{figure}

\begin{figure}
\epsfxsize=2.9in
\centerline{\epsffile{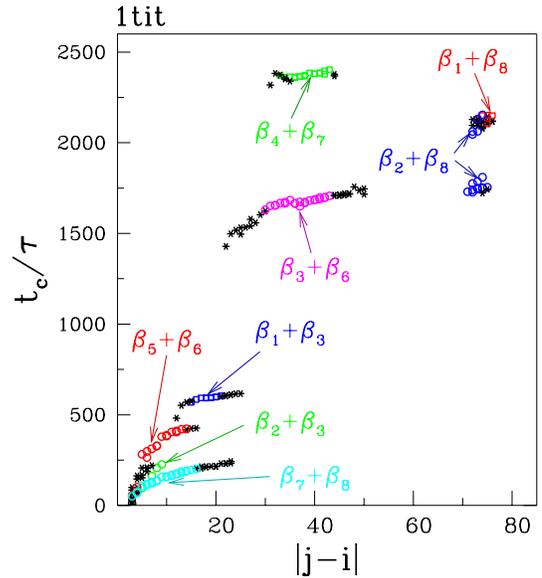}}
\vspace*{1cm}
\caption{
Sequencing of the folding events in the Go model of 1tit as represented
by the time needed to establish a contact vs. the contact order.
The open symbols correspond to contacts that form the $\beta$-sheets,
i.e. contacts between strands. The labels indicate the pairs of
strands associated with those contacts. 
The stars correspond to other contacts.
}
\end{figure}

\begin{figure}
\epsfxsize=2.5in
\centerline{\epsffile{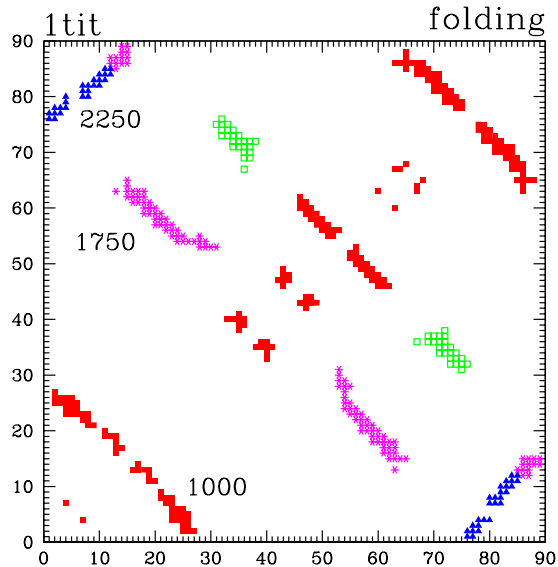}}
\vspace*{2cm}
\caption{
The contact map of 1tit without the contacts of the $i,i+2$ type.
The correspondence between $i$ and $\beta$-sheets is given in Fig. 1.
The symbols are divided into four groups
(the division is different than in Figure 4)
to illustrate the average flow of contact formation
in folding. The solid squares correspond to contacts which are established
in the first stage and are thus present at time 1000$\tau$ -- the
number shown next to the symbols. These are the short range contacts,
the turns and some antiparallel $\beta$-sheets,
which consist of members of the first two groups of Figure 4.
The stars correspond to contacts established between 1000 and 1750 $\tau$,
i.e. in the second stage of the evolution.
These are primarily the antiparallel $\beta$-sheets and some of the
longest-ranged parallel $\beta$-sheets.
The triangles show the formation of the remaining longest-ranged
parallel $\beta$-sheets in the third stage - up to the time
of 2250 $\tau$. Finally, the open squares show the intermediate
range contacts which are formed last.
}
\end{figure}

\begin{figure}
\epsfxsize=2.7in
\centerline{\epsffile{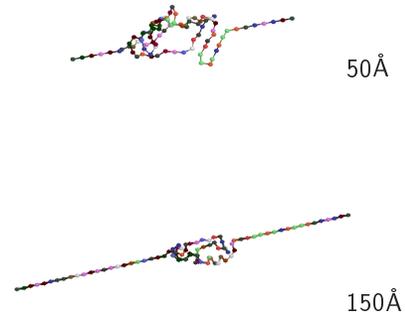}}
\caption{
Snapshot pictures of the Go model of 1tit during stretching
by a stiff cantilever at a pulling velocity of $0.005 $\AA $/\tau$.
The numbers indicate the displacement of the cantilever.
}
\end{figure}

\begin{figure}
\epsfxsize=2.3in
\centerline{\epsffile{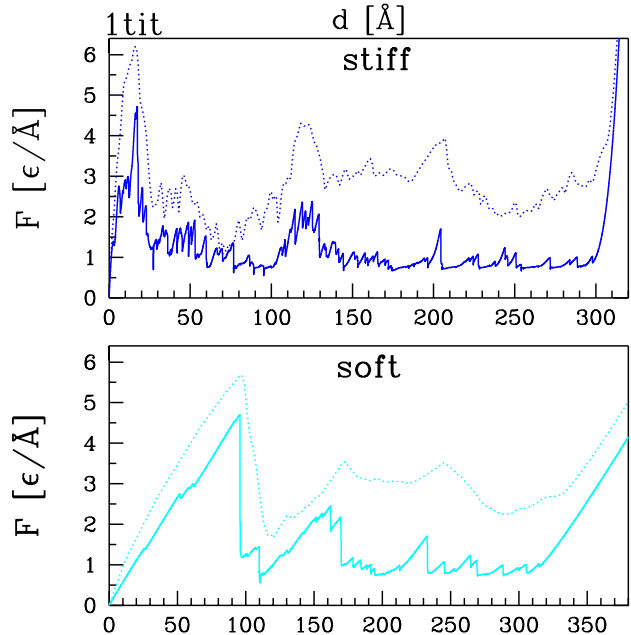}}
\vspace*{2cm}
\caption{
Force vs. cantilever displacement for the  Go model of
1tit for the slow (solid lines) and fast (dotted lines)
pulling rates.
}
\end{figure}

\begin{figure}
\epsfxsize=2.5in
\centerline{\epsffile{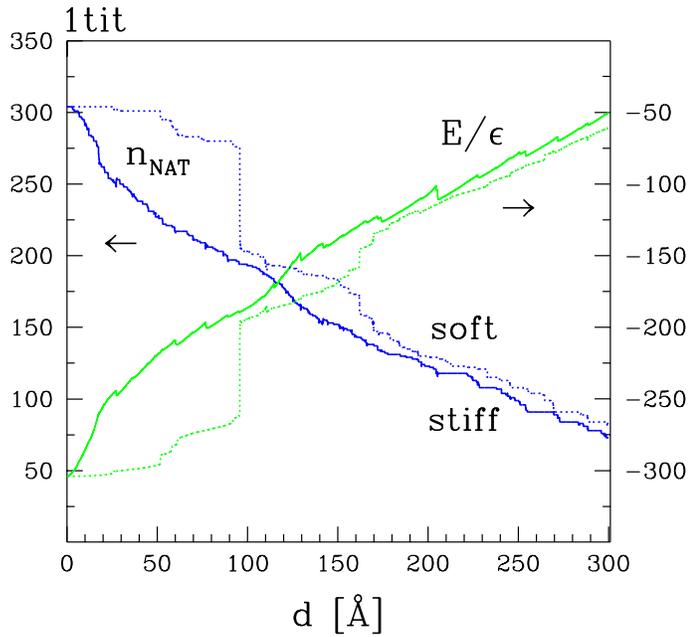}}
\vspace*{2cm}
\caption{
Energy (right axis) and number of native contacts still present
(left axis) as a function of cantilever displacement for the slow
pulling rate and for the two cantilever stiffnesses.
The number of contacts also includes those of the $i,i+2$ type.
}
\end{figure}

\begin{figure}
\epsfxsize=3.2in
\centerline{\epsffile{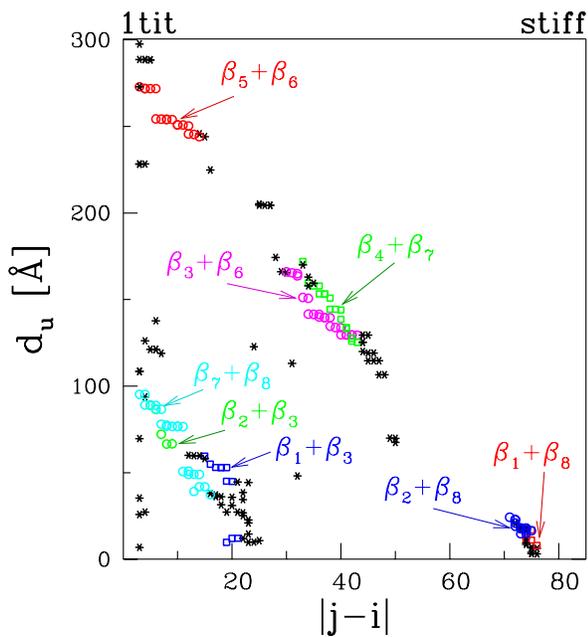}}
\vspace*{1.0cm}
\caption{
Sequencing of the stretching events in the Go model of 1tit
as represented by the cantilever displacement at which a contact breaks.
The contacts of the $i,i+2$ type are not shown here.
This is the case of the stiff cantilever which is pulled at low speed.
The symbols have the same meaning as in Figure 4.
}
\end{figure}

\begin{figure}
\epsfxsize=3.2in
\centerline{\epsffile{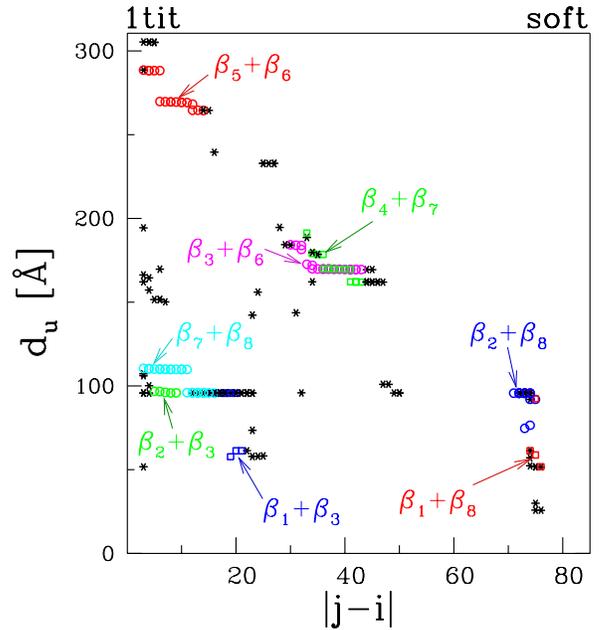}}
\vspace*{1.0cm}
\caption{
Same as in Figure 10 but for the soft cantilever.
}
\end{figure}

\begin{figure}
\epsfxsize=2.5in
\centerline{\epsffile{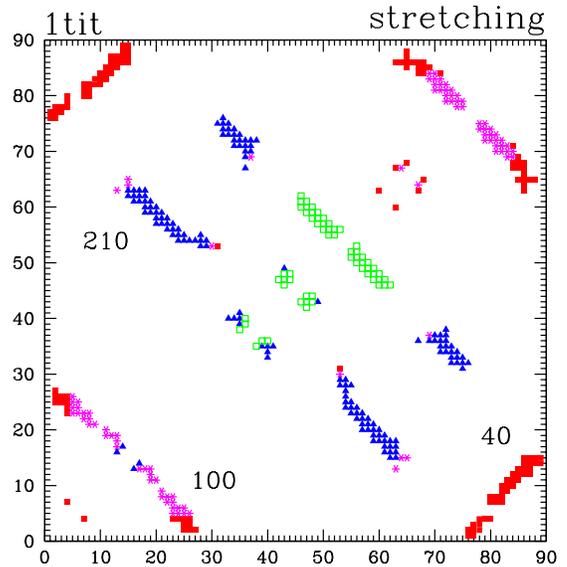}}
\vspace*{2cm}
\caption{
The contact map that represents the evolution of contact breakage
during unfolding with the stiff cantilever.
The solid squares correspond to contacts that are broken by the
time $d$ is 40 \AA,  and the stars to the additional contacts 
that are broken when $d$ is 100 \AA.
Triangles show further broken contacts when
$d$ is 210 \AA\ and open squares show contacts that break at still larger $d$.
The correspondence between $i$ and $\beta$-sheets is given in Fig. 1.
}
\end{figure}

\begin{figure}
\epsfxsize=2.5in
\centerline{\epsffile{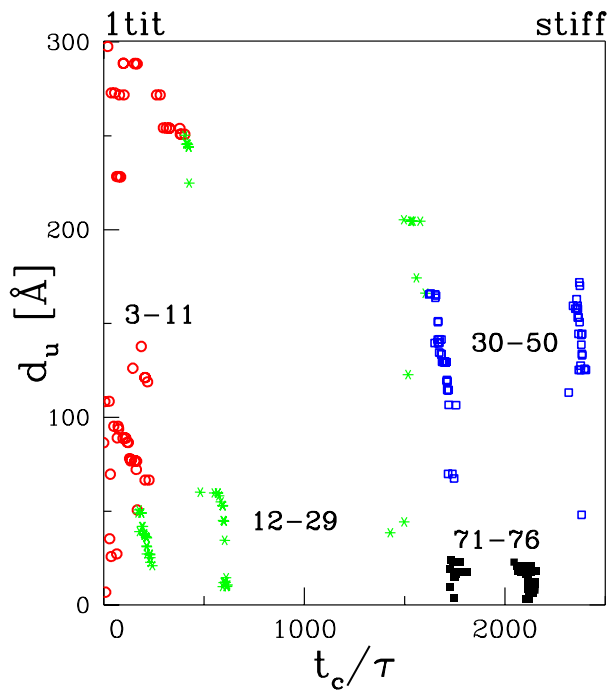}}
\vspace*{2cm}
\caption{
Stretching distances at which a bond rupture takes place
plotted vs. average time needed to establish contact on folding.
This is the case of a stiff cantilever which is being pulled slowly.
The numbers indicate the range of the contact order which 
is associated
with the symbol shown. 
}
\end{figure}

\begin{figure}
\epsfxsize=2.5in
\centerline{\epsffile{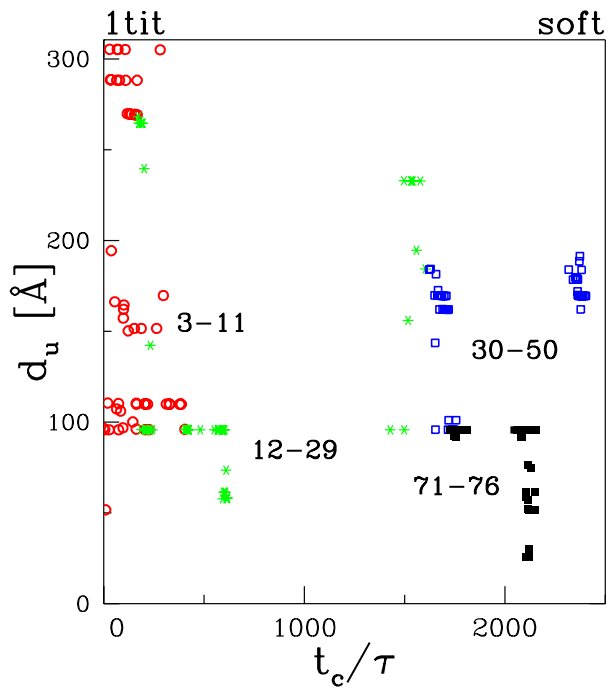}}
\vspace*{2cm}
\caption{
Similar to Figure 13 but for the case of the soft cantilever.
}
\end{figure}

\begin{figure}
\epsfxsize=2.5in
\centerline{\epsffile{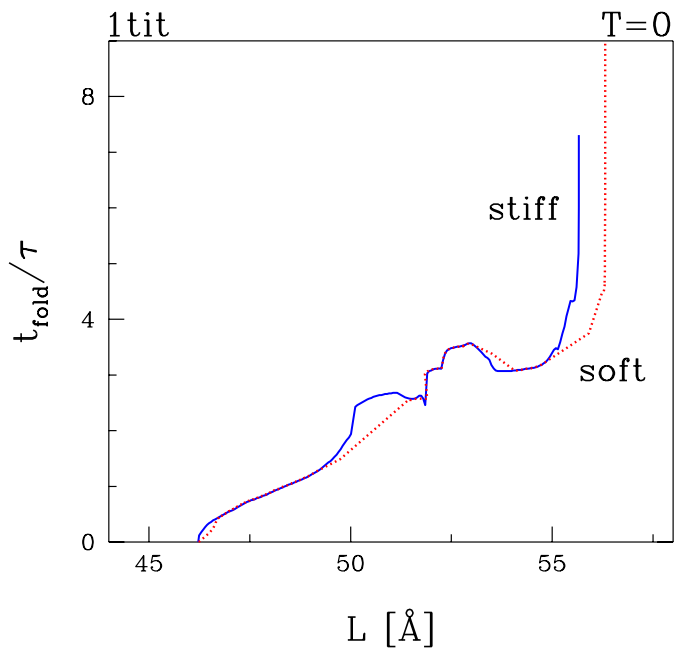}}
\vspace*{2cm}
\caption{
Refolding times for 1tit after stretching to an end-to-end distance $L$.
To the right of the data points shown,
the protein does not return to its native state but instead misfolds.
The solid lines are for a stiff cantilever and the dotted lines
are for a soft cantilever. The corresponding threshold values for the
cantilever displacement are $d_{ir}=12.6$\AA\ and $61.7$\AA for the
stiff and soft cantilevers respectively.
}
\end{figure}

\begin{figure}
\vspace*{-1cm}
\epsfxsize=2.7in
\centerline{\epsffile{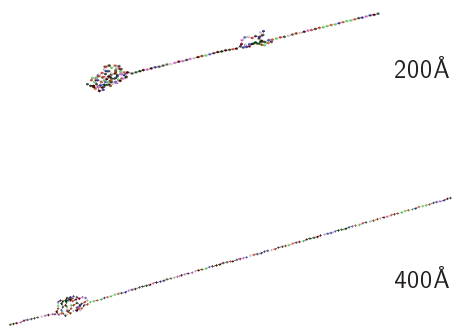}}
\vspace*{1cm}
\caption{
Snapshot pictures of the unraveling of two 1tit domains
connected in tandem. The numbers indicate the cantilever displacement.
}
\end{figure}

\begin{figure}
\epsfxsize=2.5in
\centerline{\epsffile{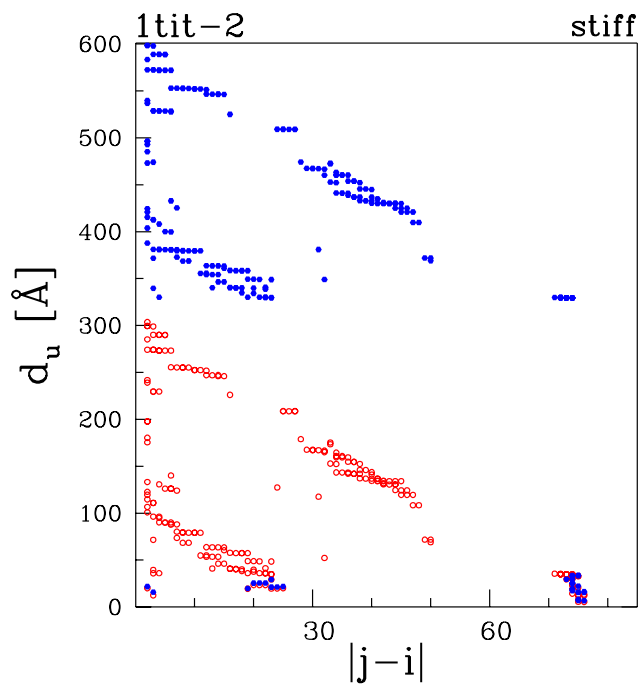}}
\vspace*{2cm}
\caption{
Sequencing of the stretching events in the Go model of two 1tit domains
as represented by $d_u$.
The contacts of the $i,i+2$ type are not shown here.
The open symbols correspond to the forward domain and the solid symbols
to the backward domain.
}
\end{figure}

\begin{figure}
\epsfxsize=2.5in
\centerline{\epsffile{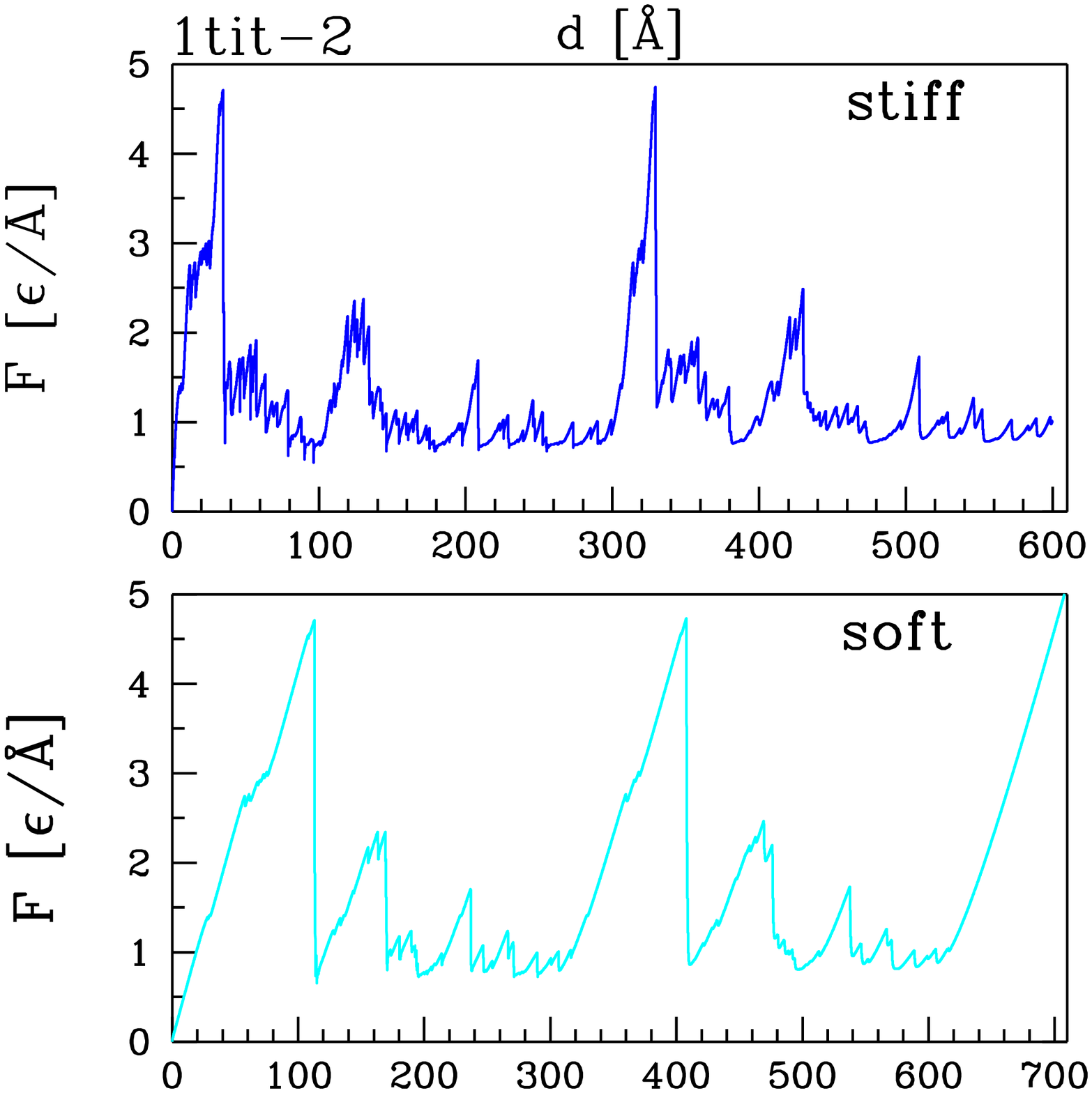}}
\vspace*{2cm}
\caption{
Force vs. cantilever displacement for two 1tit domains in series.
}
\end{figure}

\begin{figure}
\epsfxsize=2.5in
\centerline{\epsffile{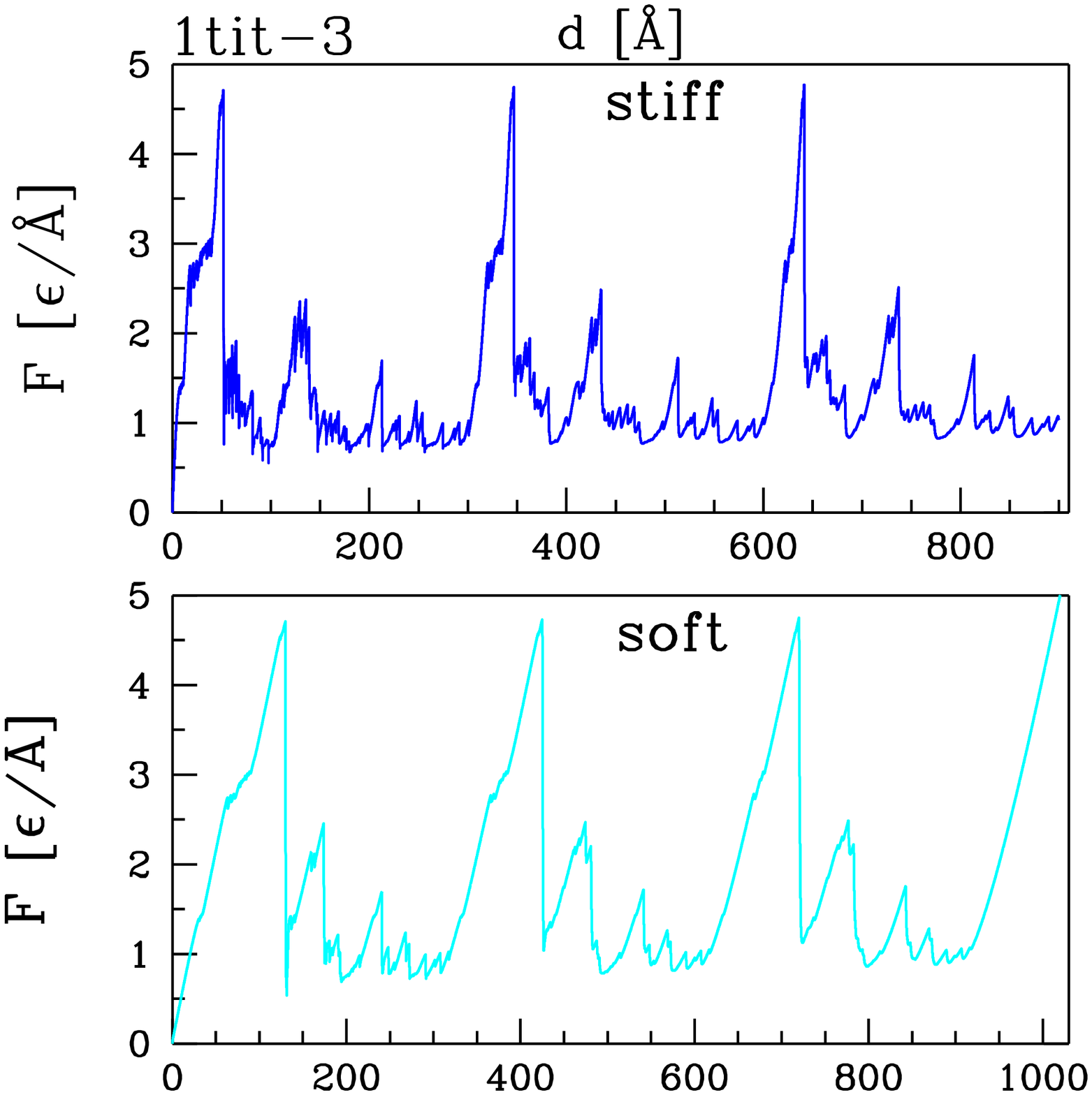}}
\vspace*{2cm}
\caption{
Force vs. cantilever displacement for three 1tit domains in series.
}
\end{figure}

\end{document}